\documentclass[sigconf,nonacm]{configs/acmart}

\makeatletter
\def\@ACM@checkaffil{
    \if@ACM@instpresent\else
    \ClassWarningNoLine{\@classname}{No institution present for an affiliation}%
    \fi
    \if@ACM@citypresent\else
    \ClassWarningNoLine{\@classname}{No city present for an affiliation}%
    \fi
    \if@ACM@countrypresent\else
        \ClassWarningNoLine{\@classname}{No country present for an affiliation}%
    \fi
}
\makeatother

\AtBeginDocument{%
  }

\IfFileExists{libertine.sty}{}{\IfFileExists{lmodern.sty}{\usepackage{lmodern}}{\usepackage{txfonts}}}
\usepackage{xcolor}
\usepackage{xspace}
\usepackage{multirow}
\usepackage{algorithm}
\usepackage[noend]{algpseudocode}
\usepackage{orcidlink}

\newcommand{\sys}{SlackSwap\xspace}
\newcommand{\keep}{\textsc{keep}\xspace}
\newcommand{\swap}{\textsc{swap}\xspace}
\newcommand{\recomp}{\textsc{recomp}\xspace}
\algrenewcommand\algorithmicindent{0.9em}

\begin{document}

\title[SlackSwap: Bandwidth-Aware Activation Swapping and Recomputation]{SlackSwap: Bandwidth-Aware Activation Swapping and Recomputation for Memory-Efficient DNN Training}

\author{Hui Feng}
\affiliation{%
  \institution{University of California Merced}
  \city{Merced}
  \state{California}
  \country{USA}
}

\author{Xun Zhou}
\affiliation{%
  \institution{University of Texas at Austin}
  \city{Austin}
  \state{Texas}
  \country{USA}
}

\author{Yuxin Wang}
\affiliation{%
  \institution{Independent Researcher}
  \city{San Francisco}
  \state{California}
  \country{USA}
}

\renewcommand{\shortauthors}{Anonymous}

\begin{abstract}
Swapping saved activations to host memory let GPUs train models larger than their memory when compute and the host link were roughly balanced: on a 2015 Titan X, PCIe could copy every VGG-16 activation out while the forward pass ran. That balance is gone. From the Titan X to the H100, peak training FLOP/s grew 141$\times$ while the host link grew 3.9$\times$, and on an A100 the link can hide only 11--38\% of the activations of the transformers we study. Swap policies such as vDNN, SuperNeurons and SwapAdvisor still push whatever the memory budget demands through the link and stall; block checkpointing leaves the link idle. We present \sys, which treats link time as a budget. From one profiled iteration, \sys spends the forward-pass link budget on the tensors that are most expensive to recompute, covers the remaining deficit by recomputing the cheapest ones, and runs FIFO offloads and earliest-deadline-first prefetches with no per-layer barriers. In a trace-driven simulation of BERT-Large, GPT-2 XL and a Llama-7B-style model on an A100, \sys keeps 96--100\% of the throughput of an unlimited-memory GPU at half the activation memory, 1.10--1.11$\times$ the best baseline, stays within 0.5\% of a genetic search over 6,600--14,000 plans, and plans in under 0.4\,s.
\end{abstract}

\ccsdesc[500]{Computer systems organization~Heterogeneous (hybrid) systems}
\ccsdesc[500]{Software and its engineering~Memory management}
\ccsdesc[300]{Computing methodologies~Machine learning}

\keywords{GPU memory management, tensor swapping, rematerialization, DNN training, PCIe}

\maketitle

\section{Introduction}

A GPT-2 XL training step over 8K tokens saves 22\,GB of activations for the backward pass, almost as much as the 25\,GB of weights, gradients and Adam states of its 1.5B parameters. When saved activations exceed device memory, a system must free them between their last forward use and their first backward use, in one of two ways. \emph{Swapping} copies a tensor to host memory over the CPU--GPU link and copies it back before it is read~\cite{rhu2016vdnn,wang2018superneurons,huang2020swapadvisor}; unified memory does this on demand, through page faults~\cite{lin2025understanding}. \emph{Recomputation} drops the tensor and re-runs the operator that produced it~\cite{chen2016sublinear,jain2020checkmate,kirisame2021dtr}; ZeRO-Offload takes this route for activations while it offloads model states~\cite{ren2021zerooffload}, and POET combines both for edge devices~\cite{patil2022poet}.

\begin{figure}[t]
  \centering
  \includegraphics[width=\columnwidth]{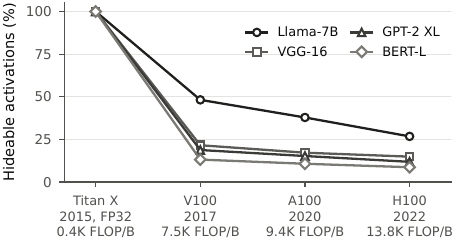}
  \caption{Share of saved activations the host link can copy out during the forward pass (Eq.~\ref{eq:frac}). The last line under each GPU gives its GEMM FLOP/s per link byte/s.}
  \label{fig:fraction}
\end{figure}

Swapping is attractive because a copy that overlaps computation is free. vDNN made this case in 2016: on a Titan X with 12.8\,GB/s of PCIe~3.0, the link could move every feature map of VGG-16 while the forward pass ran, and swapping cost 1--18\%~\cite{rhu2016vdnn}. The balance has since moved. From the Titan X (FP32) to the H100 (BF16), peak training FLOP/s grew 141$\times$ while the host link grew 3.9$\times$, from PCIe~3.0 to PCIe~5.0~\cite{nvidia2023h100}. Figure~\ref{fig:fraction} shows the result: the share of saved activations that the link can copy out during the forward pass fell from 100\% to 9--27\%.

Most swap policies still treat the link as able to absorb whatever the memory budget demands. They choose tensors by layer type~\cite{rhu2016vdnn,wang2018superneurons} or by reuse distance~\cite{huang2020swapadvisor}, and none caps the swapped bytes at what the link can hide. In our simulation of an A100 at half the activation memory, vDNN, SuperNeurons and SwapAdvisor run GPT-2 XL at 36\%, 66\% and 43\% of the throughput of a GPU with unlimited memory, because their copies sit on the critical path. Block checkpointing avoids stalls but leaves the link idle and re-runs 13\% of the step.

Two quantities decide what to do with each saved tensor. The \emph{link budget} is the number of bytes the link can copy while computation runs; copies inside it are free. The \emph{recompute cost per byte} is the critical-path time spent rebuilding a tensor, divided by its size. On an A100, rebuilding costs 1--11\,ps per byte, while an exposed PCIe~4.0 copy costs 40\,ps per byte (Figure~\ref{fig:cost}). Every byte beyond the link budget should therefore be recomputed, and the budget should go to the bytes that are most expensive to recompute.

We build this into \sys, a planner and runtime policy for activation swapping. From one profiled iteration, Pass~1 fills the forward-pass link budget with the tensors that cost most to recompute, admitting a tensor only if FIFO offloads and earliest-deadline-first (EDF) prefetches still meet their deadlines under the memory budget. Pass~2 walks the resulting memory profile and covers every over-budget point with the cheapest recomputation. At run time, two DMA queues move data without per-layer barriers, so computation waits only for a missing input or for memory. This paper makes three contributions:
\begin{itemize}
\item A link-budget model: the fraction of activations a link can hide is $\min(1, I/\beta)$, a workload's forward FLOPs per saved byte over a machine's FLOP/s per link byte/s. On an A100 it is 11--38\% for the transformers we study.
\item \sys's two-pass planner, which runs in under 0.4\,s, and its barrier-free runtime policy.
\item A trace-driven evaluation against vDNN, SuperNeurons, SwapAdvisor and block checkpointing. At half the activation memory \sys keeps 96--100\% of ideal throughput, 1.10--1.11$\times$ the best baseline; it stays within 0.5\% of a genetic search over 6,600--14,000 plans; and it trains at 10\% of the peak, where no checkpointing plan fits.
\end{itemize}

\section{Background and Motivation}

\subsection{Two ways to free a saved activation}
Training saves, for every operator, the tensors its backward pass reads. With FlashAttention~\cite{dao2022flashattention}, a GPT-style layer saves 34 bytes per token per hidden unit in BF16~\cite{korthikanti2023reducing}: the inputs of the four GEMMs (one of them the attention output), Q/K/V, the GeLU input, two LayerNorm inputs and two dropout masks. A saved tensor $t$ is produced at time $r_t$ in the forward pass and next read at $d_t$ in the backward pass. In between, it can be \emph{kept}; \emph{swapped}, copied device-to-host (D2H) after $r_t$ and host-to-device (H2D) before $d_t$ on DMA engines that run beside the compute stream; or \emph{recomputed}, dropped after its last forward reader and rebuilt before $d_t$ by re-running its producer on the compute stream.

vDNN~\cite{rhu2016vdnn} offloads the inputs of every layer (vDNN$_{all}$) or of convolutions (vDNN$_{conv}$), overlapping each copy with the layer that reads it and holding the next layer until the copy ends; in the backward pass it prefetches the closest preceding offloaded layer. SuperNeurons~\cite{wang2018superneurons} recomputes cheap layers, offloads convolution outputs through an LRU tensor cache when memory runs out, and prefetches one checkpoint ahead. SwapAdvisor~\cite{huang2020swapadvisor} evicts the tensor whose next use is furthest away~\cite{belady1966study}, issues copies as early as possible, and searches operator orders and memory-pool layouts with a genetic algorithm.

\subsection{How much can the link hide?}
A swap-out frees memory only when it completes, and the activation footprint peaks when the forward pass ends. Before the peak, the D2H direction can therefore free at most $W T_F$ bytes, where $W$ is the link bandwidth and $T_F$ the forward time. With $A$ saved bytes and $F$ forward FLOPs, the hideable fraction is
\begin{equation}
  f=\min\!\left(1,\frac{W\,T_F}{A}\right)=\min\!\left(1,\frac{I}{\beta}\right),\quad I=\frac{F}{A},\quad \beta=\frac{F/T_F}{W}.
  \label{eq:frac}
\end{equation}
$I$ belongs to the workload: forward FLOPs per saved byte. $\beta$ belongs to the machine: sustained FLOP/s per byte/s of host link. On the Titan X, $\beta\approx400$ and every workload in Figure~\ref{fig:fraction} has $I\geq403$, so the link hides everything; this is the regime vDNN was built for. Tensor cores raised $\beta$ to 4,700--6,900 on the V100 and 5,900--8,800 on the A100, while moving activations from FP32 to 16 bits only doubled $I$. GPT-2 XL has $I=1{,}200$, so an A100 can hide 15\% of its 22\,GB, and even a perfect swap-only policy stalls once the budget drops below 85\% of the peak. The H2D direction has the whole backward pass, about $2T_F$, so the D2H direction binds.

\subsection{What should happen to the rest?}
\begin{figure}[t]
  \centering
  \includegraphics[width=\columnwidth]{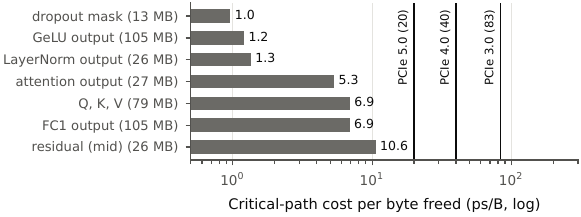}
  \caption{Critical-path time per byte to free each saved tensor of a GPT-2 XL layer on an A100 by recomputation (bars) or by an exposed copy (lines).}
  \label{fig:cost}
\end{figure}

A byte that does not fit the link budget can still be copied, with the copy exposed, or it can be recomputed. Figure~\ref{fig:cost} compares the two costs for one GPT-2 XL layer on an A100. Rebuilding costs 1.0--1.3\,ps per byte for elementwise outputs and 5.3--10.6\,ps for GEMM, attention and residual tensors; an exposed copy costs $1/W$: 83, 40 and 20\,ps per byte on PCIe~3.0, 4.0 and 5.0. Recomputation wins for every tensor, by 1.9--87$\times$. Swapping pays only inside the link budget, and there each swapped byte saves that byte's recompute cost, so the budget is best spent on tensors that are expensive or impossible to rebuild: block inputs, the residual stream and GEMM outputs. The swap policies above neither cap swapped bytes at the link budget nor rank tensors by recompute cost: they pick tensors by layer type or reuse distance and swap as many bytes as the memory budget requires. Block checkpointing~\cite{chen2016sublinear} re-runs whole blocks, paying GEMM time to rebuild elementwise bytes that cost 4--11$\times$ less to recompute alone, and never uses the link.

\section{\sys}

\subsection{Problem}
\sys takes one profiled iteration, which framework hooks or GPU analysis tools such as DrGPUM~\cite{lin2023drgpum} can collect: the operator sequence (forward, then backward), operator times, and for each saved tensor $t$ its size $s_t$, production time $r_t$, first backward use $d_t$, and recompute cost $c_t$, the time to re-run its producer from resident inputs ($c_t=\infty$ for tensors such as block inputs that no single operator rebuilds). Given a device-memory budget $M$ for activations and transient buffers, it assigns every tensor \keep, \swap or \recomp so that memory never exceeds $M$ and the iteration is as short as possible. We write $\rho_t=c_t/s_t$ for the recompute cost per byte and $P$ for the peak when every tensor is kept.

\begin{algorithm}[t]
\small
\caption{\sys planner}
\label{alg:plan}
\begin{algorithmic}[1]
\Require profile $\{s_t,r_t,d_t,c_t\}$, budget $M$, link bandwidth $W$, guard $g$
\State $x_t \gets$ \keep for all $t$;\; $D \gets P-M$;\; \textbf{if} $D\le 0$ \textbf{return} $x$
\State $S\gets\emptyset$ \Comment{Pass 1: spend the link budget}
\For{$t$ with $r_t<T_F<d_t$, by decreasing $\rho_t$}
  \State \textbf{if} $\sum_{u\in S}s_u\ge D$ \textbf{break}
  \If{\textsc{FifoD2H}$(S\cup\{t\})$ ends before $(1-g)T_F$ \textbf{and} \textsc{EdfH2D}$(S\cup\{t\})$ meets every $d_u$}
    \State $S\gets S\cup\{t\}$;\; $x_t\gets$ \swap
  \EndIf
\EndFor
\Loop \Comment{Pass 2: cover the rest with recomputation}
  \State $m\gets$ \textsc{MemoryProfile}$(x)$;\; $i\gets$ first op with $m(i)>M$
  \State \textbf{if} no such $i$ \textbf{return} $x$
  \State $E\gets\max m-M$;\; $C\gets\{t:$ resident at $i$, unused until after $i$, $c_t<\infty\}$
  \If{$C=\emptyset$} convert the earliest remaining block to full recomputation
  \Else\; $x_{t^*}\gets$ \recomp for $t^*=\arg\min_{t\in C}\, c_t/\min(s_t,E)$
  \EndIf
\EndLoop
\end{algorithmic}
\end{algorithm}

\subsection{Pass 1: spend the link budget}
Pass~1 (Algorithm~\ref{alg:plan}, lines 2--6) considers tensors produced in the forward pass and read in the backward pass in decreasing $\rho_t$: block inputs ($\rho=\infty$) first, then the residual stream and GEMM and attention outputs, then elementwise outputs. It adds $t$ to the swap set $S$ only if two schedules stay feasible. \emph{Offload}: D2H copies run in production order, each starting no earlier than its $r_t$, and all end before $(1-g)T_F$, with a 2\% guard $g$. \emph{Prefetch}: H2D copies start at $T_F$ and run in deadline order, which minimizes the maximum lateness on one channel~\cite{jackson1955scheduling}. A copy may start only once the backward pass has released enough memory for it: we assume memory is full at $T_F$ and credit the bytes of each kept tensor when its last backward reader ends. Every copy must end before its $d_u$. Pass~1 stops when $S$ covers the deficit $D=P-M$ or no candidate passes; each check takes $O(|S|\log|S|)$ time.

\subsection{Pass 2: recompute the cheapest bytes}
Pass~2 (lines 7--12) computes the memory profile $m(i)$ over the operator sequence. A \keep tensor occupies memory from production to its last use; a \swap tensor until its offload ends (and its last forward reader is done) and again from its first backward use; a \recomp tensor until its last forward reader and again once rebuilt. At the first operator $i$ where $m(i)>M$, Pass~2 looks at tensors that are resident at $i$ but unused until after it, and marks \recomp the one with the lowest cost per byte of the excess it can cover, $c_t/\min(s_t,E)$, where $E$ is the largest remaining excess. When tensors are small relative to $E$ this is the fractional-knapsack rule, lowest $\rho_t$ first; when they are not, it avoids rebuilding a large tensor where a small one suffices. Rebuilds may chain: if a \recomp tensor's input is also \recomp, the input is rebuilt first. If no candidate is left, Pass~2 converts the earliest remaining block to full recomputation, i.e., block checkpointing as a fallback. Because swapped block inputs also leave memory, \sys can go below the floor of checkpointing alone.

\begin{figure}[t]
  \centering
  \includegraphics[width=\columnwidth]{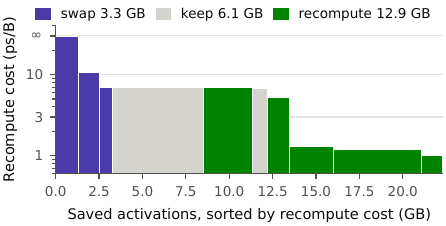}
  \caption{\sys's plan for GPT-2 XL at 30\% of the peak on an A100. Each bar is a group of saved tensors: width is size, height is recompute cost per byte.}
  \label{fig:plan}
\end{figure}

Figure~\ref{fig:plan} shows the resulting plan for GPT-2 XL at 30\% of its peak. Block inputs, mid-block residual tensors and the Q/K/V of the first nine layers fill the 3.3\,GB link budget; every elementwise output, 46 of 48 attention outputs and the FC1 outputs of the first 27 layers are recomputed; the remaining GEMM outputs stay resident.

\subsection{Validation, repair and runtime}
\paragraph{Validation and repair.} The memory profile ignores second-order effects: prefetches that wait for memory, rebuild chains, and transient buffers at the forward/backward boundary. \sys therefore replays the plan once in an event model of the iteration (in deployment, one dry-run step traced with a tool such as PASTA~\cite{lin2026pasta}) and repairs what it observes. A tensor whose prefetch held up the compute stream is recomputed if that is cheaper than the wait, or else removed from $S$ before re-planning; lagging offloads raise $g$; a plan that does not fit is re-planned for a budget 1\%, 2\%, 4\%,\,\dots\ smaller. \sys keeps the fastest plan seen in at most six rounds, including the plan with no swaps.

\paragraph{Runtime.} The runtime uses two copy streams. A D2H copy starts as soon as its tensor is produced; saved tensors are read-only, so the copy overlaps the operators that still read it, and the device buffer is released once both the copy and the last forward reader are done. When the backward pass starts, H2D copies are queued in deadline order, and each is admitted when used memory, its size and the allocations of the next three operators fit in $M$. No operator waits on a copy it does not need: there is no per-layer barrier. If the compute stream lacks memory while no copy is in flight, the runtime frees device copies of swapped tensors that already have a host copy, furthest next use first, and queues them again. Rebuild operators run immediately before their first backward reader. In PyTorch, these mechanisms map onto saved-tensor pack/unpack hooks, pinned host buffers and two CUDA streams; this paper evaluates the policy in simulation.

\section{Evaluation}

\subsection{Methodology}
\paragraph{Simulator.} A discrete-event simulator models one training iteration on one GPU: a compute stream, a D2H and an H2D DMA engine (each at the link bandwidth, 85\% of it when both are busy, plus 10\,$\mu$s per copy), and a memory budget covering saved activations, rebuilt tensors, prefetch buffers, gradient buffers and temporaries. Operator times follow a roofline with datasheet peaks~\cite{nvidia2021a100,nvidia2023h100} and sustained efficiencies: 75\% of 312\,TFLOPS for A100 GEMMs, 62\%/48\% for FlashAttention forward/backward, 85\% of 2.0\,TB/s for memory-bound kernels, and 5\,$\mu$s per launch. The link is PCIe~4.0 at 25\,GB/s per direction. Model states stay on the GPU, and budgets are fractions of each workload's no-swap peak $P$. The simulator assumes a fragmentation-free allocator and ignores the 1.2\% of HBM bandwidth that copies take. As a check, with vDNN's Titan X parameters (7\,TFLOPS FP32, 12.8\,GB/s) it predicts 3.4\% overhead for vDNN$_{conv}$ and 22\% for vDNN$_{all}$ on VGG-16; vDNN reports 1--18\% for its dynamic policy~\cite{rhu2016vdnn}, which uses vDNN$_{conv}$ when it fits.

\paragraph{Workloads.} Table~\ref{tab:workloads} lists BERT-Large~\cite{devlin2019bert}, GPT-2 XL~\cite{radford2019gpt2}, a Llama-2-7B-style model (RMSNorm, SwiGLU, no dropout)~\cite{touvron2023llama2} and VGG-16~\cite{simonyan2015vgg}, all in BF16 with FlashAttention for the transformers.

\paragraph{Baselines.} All run in the same simulator. \emph{vDNN}: the dynamic policy (no offload if it fits, else vDNN$_{conv}$, else vDNN$_{all}$) with per-layer barriers and one-layer-ahead prefetch. \emph{SuperNeurons}: recomputes LayerNorm, GeLU and dropout outputs, evicts GEMM and attention outputs on demand in LRU order, and prefetches for the next heavy operator. \emph{SwapAdvisor}: Belady eviction on the planned schedule with as-early-as-possible swaps; we omit its genetic search over operator order and memory pools, which has little room on these near-linear graphs with our allocator, as its authors observe for WResNet. \emph{Checkpoint}: the fewest blocks recomputed from their inputs that fit, as in ZeRO-Offload and Megatron-LM~\cite{shoeybi2019megatron}.

\begin{table}[t]
  \caption{Workloads on the A100: saved activations $A$, ideal step time, intensity $I$ and hideable fraction $f$ (Eq.~\ref{eq:frac}).}
  \label{tab:workloads}
  \footnotesize
  \setlength{\tabcolsep}{3pt}
  \begin{tabular}{@{}llrrrr@{}}
    \toprule
    Model & Configuration & $A$ (GB) & Step (ms) & $I$ (F/B) & $f$ \\
    \midrule
    BERT-Large & 24L, $h$=1024, $s$=512, $b$=32 & 13.9 & 173 & 782 & 11\% \\
GPT-2 XL & 48L, $h$=1600, $s$=1024, $b$=8 & 22.3 & 399 & 1,200 & 15\% \\
Llama-7B & 32L, $h$=4096, $s$=4096, $b$=2 & 35.2 & 1,635 & 3,326 & 38\% \\
VGG-16 & $b$=128, 224$\times$224 & 3.9 & 82 & 1,015 & 17\% \\
    \bottomrule
  \end{tabular}
\end{table}

\begin{figure*}[t]
  \centering
  \includegraphics[width=\textwidth]{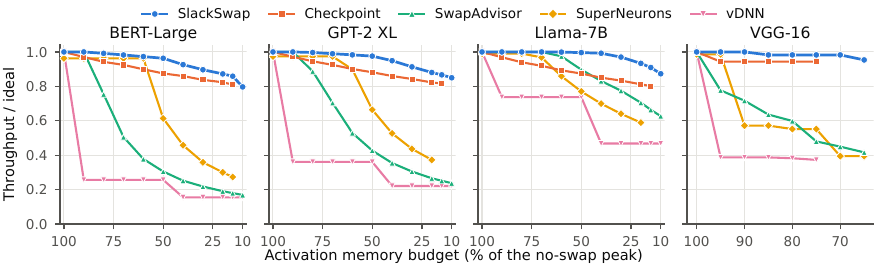}
  \caption{Throughput relative to a GPU with unlimited memory as the activation budget shrinks (A100, PCIe 4.0). A missing point means the policy cannot fit the budget.}
  \label{fig:main}
\end{figure*}

\subsection{Throughput across memory budgets}
Figure~\ref{fig:main} reports throughput relative to the ideal. \sys is the top curve, or tied for it, at every budget. Down to 90\% of the peak (70\% for Llama-7B), the deficit fits in the link budget, and \sys and SwapAdvisor both match the ideal. Below that, SwapAdvisor pushes more bytes than the link moves: at 50\% it copies 11.6\,GB out for GPT-2 XL during a 136\,ms forward pass and falls to 43\%. Recomputing cheap layers carries SuperNeurons down to 60--70\% of the peak at 96--99\%, after which on-demand eviction puts every copy on the critical path. vDNN's barriers expose its copies even when only a few are needed; it drops to 26--74\% as soon as any offload starts. Checkpointing degrades smoothly to 80--82\% at 15\% of the peak and cannot go lower.

At 50\% of the peak, \sys keeps 96.3\%, 97.6\% and 99.6\% of the ideal for BERT-Large, GPT-2 XL and Llama-7B, 1.10--1.11$\times$ the best baseline; at 30\% it keeps 89.7--97.0\%, 1.07--1.16$\times$. Llama-7B gains most because its link budget is the largest (38\%). At 10\%, where no checkpointing plan fits, \sys still runs at 79.7--87.3\% by swapping block inputs. VGG-16 keeps 42\% of its activations in its first two convolutions, so its feasible range is narrow; \sys stays at 95--100\% down to 65\% of the peak, while vDNN and checkpointing stop fitting at 70\%.

\begin{table}[t]
  \caption{Where the time goes at 40\% of the peak: step time, recomputation and stalls as multiples of the ideal step, and GB copied out.}
  \label{tab:breakdown}
  \footnotesize
  \setlength{\tabcolsep}{3.2pt}
  \begin{tabular}{@{}llrrrrr@{}}
    \toprule
    & Policy & Step & Recomp. & Fwd stall & Bwd stall & D2H GB \\
    \midrule
    \multirow{5}{*}{GPT-2 XL} & vDNN & 4.52 & 0.000 & 1.85 & 1.67 & 22.3 \\
 & SuperNeurons & 1.90 & 0.027 & 0.43 & 0.44 & 5.1 \\
 & SwapAdvisor & 2.82 & 0.000 & 0.99 & 0.83 & 14.0 \\
 & Checkpoint & 1.16 & 0.163 & 0.00 & 0.00 & 0.0 \\
 & \textbf{SlackSwap} & 1.05 & 0.053 & 0.00 & 0.00 & 3.3 \\
\midrule
\multirow{5}{*}{Llama-7B} & vDNN & 2.14 & 0.000 & 0.62 & 0.52 & 35.2 \\
 & SuperNeurons & 1.43 & 0.009 & 0.27 & 0.15 & 11.3 \\
 & SwapAdvisor & 1.20 & 0.000 & 0.19 & 0.01 & 21.6 \\
 & Checkpoint & 1.18 & 0.176 & 0.00 & 0.00 & 0.0 \\
 & \textbf{SlackSwap} & 1.01 & 0.008 & 0.00 & 0.00 & 13.0 \\
    \bottomrule
  \end{tabular}
\end{table}

\subsection{Where the time goes}
\begin{figure}[t]
  \centering
  \includegraphics[width=\columnwidth]{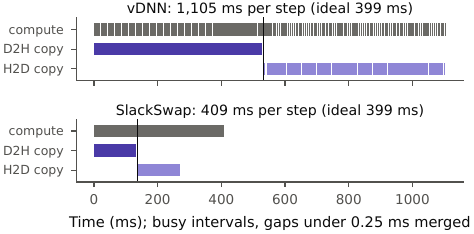}
  \caption{One GPT-2 XL step at 50\% of the peak on an A100. The vertical line ends the forward pass; white gaps in the compute lane are stalls.}
  \label{fig:gantt}
\end{figure}

Table~\ref{tab:breakdown} splits the step at 40\% of the peak. Swap policies lose time to exposed copies: SwapAdvisor's GPT-2 XL step spends 0.99 and 0.83 ideal steps stalled in the forward and backward passes. Checkpointing loses 16--18\% to recomputation. \sys copies out 3.3\,GB for GPT-2 XL and 13.0\,GB for Llama-7B, the whole forward link budget, recomputes 5.3\% and 0.8\% of the step, and stalls for less than 0.1\%. Figure~\ref{fig:gantt} shows the GPT-2 XL step at 50\%: vDNN's barriers stretch it to 1,105\,ms, while \sys's D2H queue is busy for the whole forward pass, its H2D queue for the first part of the backward pass, and the step takes 409\,ms against 399\,ms ideal.

\subsection{Ablation and optimality}
Table~\ref{tab:ablation} removes one idea at a time. \emph{Swap-first} ranks swap candidates by production time, the Belady-like order, instead of recompute cost: it spends part of the link budget on cheap elementwise bytes and then pays for GEMM recomputation, losing 0.1--7.9 points. \emph{Recompute-only} drops Pass~1 and loses 3.0--10.7 points, the value of the link. \emph{Barriers} runs \sys's plan with vDNN's per-layer synchronization and loses 32--37 points: the plan pays off only with a barrier-free runtime.

To bound the remaining gap, we run a genetic search over per-tensor \keep/\swap/\recomp decisions, POET's search space with SwapAdvisor's search method, using the simulator as fitness for 45\,s (6,600--14,000 plans). Seeded with \sys and the baselines, it improves on \sys by at most 0.45\%. Started from random plans, as SwapAdvisor's search is, it finds no feasible plan at 20\% within 45\,s and trails \sys at 50\%. \sys plans in 67--168\,ms here and in at most 376\,ms across all of Figure~\ref{fig:main}, in Python.

\begin{table}[t]
  \caption{Throughput (\% of ideal) of \sys, its ablations and a 45\,s genetic search (GA), and \sys's planning time.}
  \label{tab:ablation}
  \footnotesize
  \setlength{\tabcolsep}{2.4pt}
  \begin{tabular}{@{}lrrrrrrrr@{}}
    \toprule
    & & & Swap- & Recomp. & Bar- & GA & GA & Plan \\
    Model & $M$ & \sys & first & only & riers & seeded & random & ms \\
    \midrule
    BERT-Large & 50\% & \textbf{96.3} & 94.7 & 93.1 & 59.1 & 96.4 & 94.5 & 67 \\
BERT-Large & 20\% & \textbf{87.2} & 85.4 & 84.2 & 55.5 & 87.3 & none & 83 \\
GPT-2 XL & 50\% & \textbf{97.6} & 96.6 & 93.5 & 62.2 & 97.6 & 95.1 & 119 \\
GPT-2 XL & 20\% & \textbf{88.1} & 85.7 & 84.1 & 56.3 & 88.3 & none & 168 \\
Llama-7B & 50\% & \textbf{99.6} & 99.5 & 91.5 & 66.9 & 99.6 & 99.6 & 78 \\
Llama-7B & 20\% & \textbf{93.4} & 85.5 & 82.7 & 61.7 & 93.8 & none & 95 \\
    \bottomrule
  \end{tabular}
\end{table}

\subsection{Sensitivity}
\begin{figure}[t]
  \centering
  \includegraphics[width=\columnwidth]{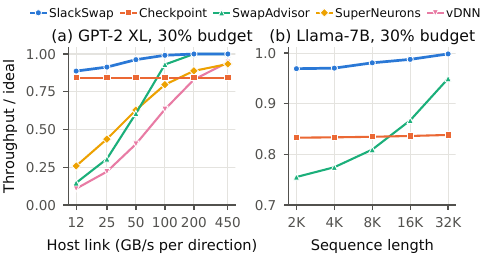}
  \caption{Sensitivity at 30\% of the peak: (a) host-link bandwidth, (b) sequence length (note the y-axis starts at 0.7).}
  \label{fig:sens}
\end{figure}

\paragraph{Link bandwidth.} Figure~\ref{fig:sens}a sweeps the link for GPT-2 XL at 30\% of the peak. \sys moves from 1.6\,GB swapped and 14.6\,GB recomputed at 12\,GB/s to 16.7\,GB swapped and nothing recomputed at 200\,GB/s, staying 0--12 points above the best baseline. Swap-centric policies need 100--450\,GB/s per direction, 4--18$\times$ PCIe~4.0, before they beat checkpointing.

\paragraph{Sequence length.} Attention FLOPs grow with sequence length while saved bytes per token do not, so $I$ and $f$ grow: for the Llama-7B model, $f$ rises from 36\% at 2K tokens to 60\% at 32K (Figure~\ref{fig:sens}b, 16K tokens per batch, 32K at the last point). \sys approaches the ideal (96.9\% to 99.8\%) while checkpointing stays at 83--84\%.

\paragraph{GPU generation.} At 30\% of the peak, \sys's lead over the best baseline is 1.10$\times$, 1.08$\times$ and 1.07$\times$ for GPT-2 XL on the V100, A100 and H100, and 1.15$\times$, 1.16$\times$ and 1.11$\times$ for Llama-7B. The H100's smaller link budget shifts \sys toward recomputation instead of stalls.

\paragraph{Limitations.} All results are simulated. The simulator does not model allocator fragmentation, HBM contention from copies, or GPUs that share a PCIe switch or host memory bandwidth; sharing shrinks each GPU's link budget, which \sys absorbs if the profile measures the shared bandwidth. We evaluate one GPU with resident model states; combining \sys with ZeRO-Offload's gradient traffic, which uses D2H during the backward pass while activation swapping uses it during the forward pass, is future work, as is applying the link budget to unified-memory prefetching~\cite{lin2026autouvm}.

\section{Related Work}
\paragraph{Swapping.} vDNN~\cite{rhu2016vdnn}, TFLMS~\cite{le2018tflms}, SuperNeurons~\cite{wang2018superneurons} and SwapAdvisor~\cite{huang2020swapadvisor} offload activations (and, for SwapAdvisor, parameters) to host memory; FlashNeuron~\cite{bae2021flashneuron} offloads to SSDs and Sentinel~\cite{ren2021sentinel} places tensors across heterogeneous memory. Demand paging through unified memory is slower than explicit copies~\cite{zheng2016paged,rhu2016vdnn}; recent work characterizes oversubscribed memory management for deep learning training~\cite{lin2025understanding} and improves UVM management and prefetching~\cite{lin2025forest,lin2026autouvm}. ZeRO-Offload and ZeRO-Infinity~\cite{ren2021zerooffload,rajbhandari2021zeroinfinity} offload model states, and HybridGen~\cite{lin2026hybridgen} splits LLM inference across the CPU and GPU. The swapping systems above size swaps by memory pressure; \sys sizes them by link time.

\paragraph{Recomputation.} Chen et al.~\cite{chen2016sublinear} checkpoint every $\sqrt{n}$ layers; Checkmate~\cite{jain2020checkmate} finds optimal schedules with integer programming; DTR~\cite{kirisame2021dtr} evicts greedily at run time; Megatron-LM recomputes selected attention operators~\cite{korthikanti2023reducing}.

\paragraph{Hybrids.} Capuchin~\cite{peng2020capuchin} chooses between swapping and recomputation per tensor from access patterns observed at run time; POFO~\cite{beaumont2021pofo} combines them optimally for chains with dynamic programming; POET~\cite{patil2022poet} solves a mixed-integer program for energy on edge devices. \sys is a closed-form, millisecond planner built around one observation: on compute-rich GPUs the link budget is small and explicit, and it belongs to the tensors that are most expensive to rebuild.

\section{Conclusion}
GPU compute has outgrown the host link by more than an order of magnitude since swapping was proposed, and an A100 can now hide only 11--38\% of a transformer's saved activations. \sys plans for this directly: it spends the link budget on the tensors that are most expensive to recompute and recomputes the rest, and its runtime keeps every copy off the critical path. In simulation it holds 96--100\% of ideal throughput at half the activation memory and reaches budgets that block checkpointing cannot. The same analysis says when swapping becomes the main tool again: with coherent links of hundreds of GB/s or with long sequences, \sys shifts on its own from recomputation to swapping.

\bibliographystyle{configs/ACM-Reference-Format-citeorder}
\bibliography{references}

\end{document}